\documentclass[fleqn,usenatbib]{rasti}

\usepackage{newtxtext,newtxmath}
\usepackage[T1]{fontenc}

\DeclareRobustCommand{\VAN}[3]{#2}
\let\VANthebibliography\thebibliography
\def\thebibliography{\DeclareRobustCommand{\VAN}[3]{##3}\VANthebibliography}

\usepackage{graphicx}	% Including figure files
\usepackage{aas_macros}
\usepackage{siunitx}
\usepackage{booktabs}
\usepackage{subfig}
\DeclareSIUnit{\sqrthz}{\ensuremath{\sqrt{\text{\hertz}}}}
\DeclareSIUnit{\vrms}{\volt_{RMS}}
\DeclareSIUnit{\mps}{(ms^{-1})}

\title[BiSON:NG Instrument]{The next generation Birmingham Solar Oscillations Network (BiSON) spectrophotometer: a new miniaturised instrument for helioseismology}

\author[S. J. Hale et al.]{
S. J. Hale,$^{1}$\thanks{E-mail: s.j.hale@bham.ac.uk}
W. J. Chaplin,$^{1}$
G. R. Davies,$^{1}$
Y. P. Elsworth,$^{1}$
R. Howe,$^{1}$
\\
$^{1}$School of Physics and Astronomy, University of Birmingham, Edgbaston, Birmingham B15 2TT, United Kingdom\\
}

\date{Accepted XXX. Received YYY; in original form ZZZ}

\pubyear{2022}

\begin{document}
\label{firstpage}
\pagerange{\pageref{firstpage}--\pageref{lastpage}}
\maketitle

% -*- coding: utf-8 -*-
%
% ABSTRACT.TEX
%
%   Steven Hale
%   2022 February 11
%   Birmingham, UK
%
% Abstract
%

%%%%%%%%%%%%%%%%%%%%%%%%%%%%%%%%%%%%%%%%%%%%%%%%%%%%%%%%%%%%%%%%%%%%%%%%%%%%%%%%

% Abstract of the paper
\begin{abstract}
We describe a new spectrophotometer for the Birmingham Solar
Oscillations Network (BiSON), based on a next generation observation
platform, BiSON:NG, a significantly miniaturised system making use of
inexpensive consumer-grade hardware and off-the-shelf components,
where possible.  We show through system modelling and simulation,
along with a summer observing campaign, that the prototype instrument
produces data on the Sun's low-degree acoustic (p-mode) oscillations
that are of equal quality and can be seamlessly integrated into the
existing network.  Refreshing the existing ageing hardware, and the
extended observational network potential of BiSON:NG, will secure our
ongoing programme of high-quality synoptic observations of the Sun's
low-degree oscillations (e.g.,~for seismic monitoring of the solar
cycle at a ``whole Sun'' level).
\end{abstract}

% Include between one and six keywords.

%%%%%%%%%%%%%%%%%%%%%%%%%%%%%%%%%%%%%%%%%%%%%%%%%%%%%%%%%%%%%%%%%%%%%%%%%%%%%%%%

%%%%%%%%%%%%%%%%% BODY OF PAPER %%%%%%%%%%%%%%%%%%

% -*- coding: utf-8 -*-
%
% INTRODUCTION.TEX
%
%   Steven Hale
%   2022 February 11
%   Birmingham, UK
%
% Introduction to BiSON:NG spectrophotometer design.
%

%%%%%%%%%%%%%%%%%%%%%%%%%%%%%%%%%%%%%%%%%%%%%%%%%%%%%%%%%%%%%%%%%%%%%%%%%%%%%%%%

\section{Introduction}
\label{s:introduction}

The Birmingham Solar Oscillations Network (BiSON) observes acoustic
oscillations of the Sun-as-a-star, via a network of ground-based
automated telescopes~\citep{brookesphd,1978MNRAS.185....1B,2016SoPh..291....1H,halephd}.
The observed modes are of low angular degree, and provide data for
characterising the solar activity cycle at the ``whole Sun'' level
(see,~e.g.,~\citealt{2019MNRAS.489L..86C,10.1093/mnras/stac1534}), and
the structure and dynamics of the deep solar interior.  The network is
composed of six sites -- Mount Wilson, Los~Angeles, USA; Las Campanas,
Chile; Iza\~na, Tenerife, Canary Islands; Sutherland, South Africa;
Carnarvon, Western Australia; and Narrabri, New South Wales,
Australia.  Four of the sites are fully automated and use a 4~metre
observatory dome and a large equatorial mount.  Two of the sites,
Mount Wilson and Iza\~na, collect light via a c{\oe}lostat and so
require observers.  In this paper, we describe a new instrument based
on a next generation observation platform,
BiSON:NG~\citep{10.1117/12.2561282}, a significantly miniaturised
system making use of inexpensive consumer-grade hardware and
off-the-shelf components~\citep{halephd}.

The miniaturised system separates the collection optics from the
instrument via an optical fibre, and so requires only a small amateur
telescope mount and small enclosure, which can be inexpensively
installed on the roof of existing infrastructure.  Moving the
instrumentation off a mount and away from direct sunlight improves
thermal stability and robustness of the data.  Use of off-the-shelf
hardware, where possible, simplifies the design, operation, and
reduces cost in comparison with bespoke construction.  Ensuring that
several sources of similar or ideally identical components exists
removes a supplier single-point failure, and helps maintenance
requirements particularly on long term projects.

In Section~\ref{s:rss} we give an overview of the operation of the
instrument, and model the expected performance.  In
Section~\ref{s:noise} we analyse sources of noise that contribute to
the expected overall white noise background level, and in
Section~\ref{s:acquisition} describe how data are acquired and
recorded.  Finally, in Section~\ref{s:commissioning}, we show initial
results from a prototype trial during 2018 at Iza\~na.

%%%%%%%%%%%%%%%%%%%%%%%%%%%%%%%%%%%%%%%%%%%%%%%%%%%%%%%%%%%%%%%%%%%%%%%%%%%%%%%%

% -*- coding: utf-8 -*-
%
% RSS.TEX
%
%   Steven Hale
%   2022 February 11
%   Birmingham, UK
%
% A new resonance scattering spectrophotometer.
%

%%%%%%%%%%%%%%%%%%%%%%%%%%%%%%%%%%%%%%%%%%%%%%%%%%%%%%%%%%%%%%%%%%%%%%%%%%%%%%%%

\section{A Miniature Spectrophotometer}
\label{s:rss}

\begin{figure*}
  \includegraphics[width=0.8\textwidth]{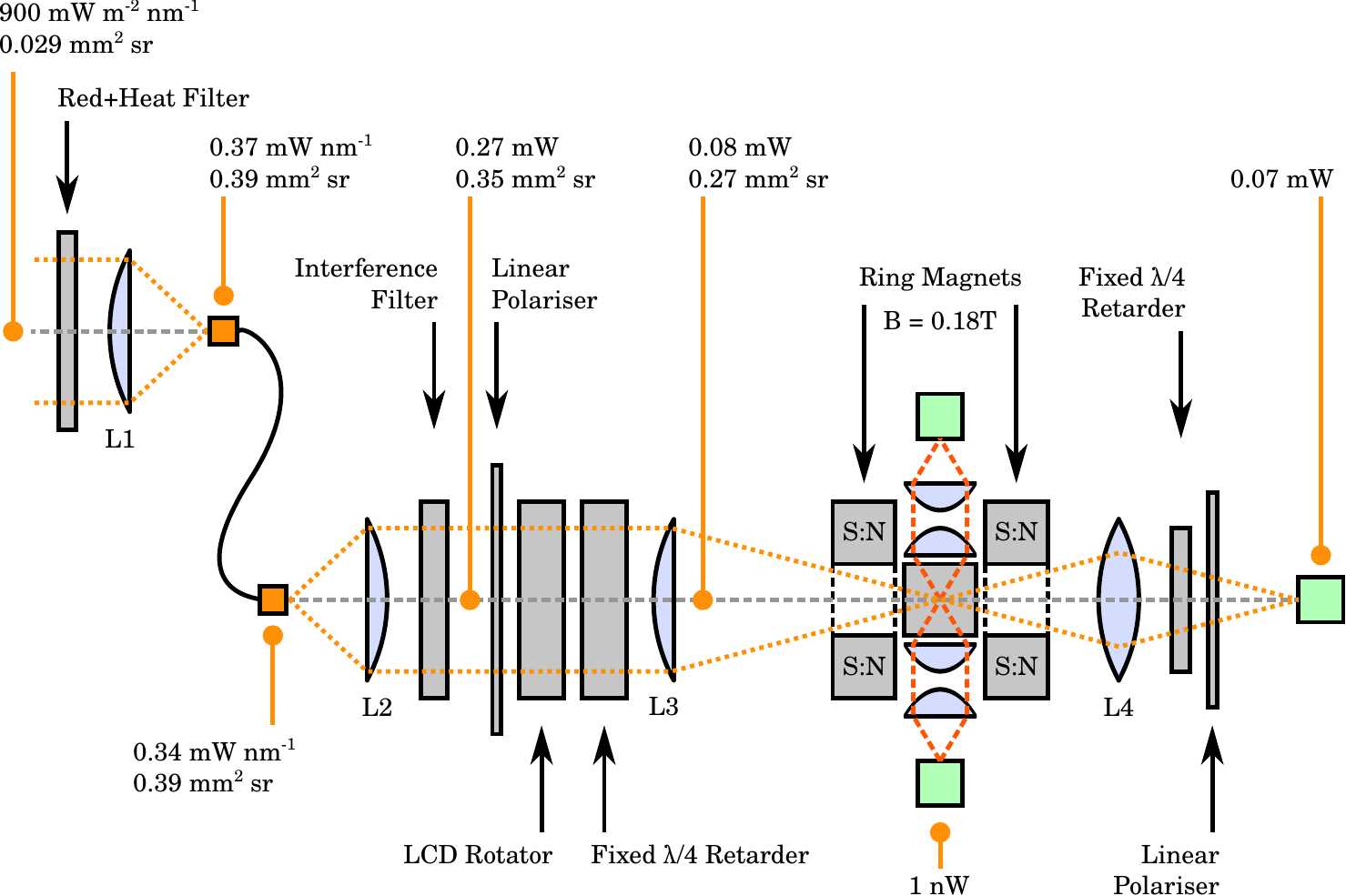}
  \caption{Schematic of a fibre-based BiSON resonance scattering
    spectrophotometer.  The {\'e}tendue and optical power (or power
    spectral density, as appropriate) is indicated at key points in
    the system by the orange markers.  The green boxes indicate
    detector locations - two perpendicular to the optical axis
    detecting light scattered from the vapour cell, and one detecting
    the light transmitted directly through the cell.}
  \label{fig:fibre_rss}
\end{figure*}

All BiSON instruments are resonance scattering spectrophotometers
(RSS).  Light scattered from a potassium vapour cell is used to make
high-precision photometric measurements of the Doppler velocity of the
solar surface, by observing fluctuations in the intensity of the
D1~transition at~\SI{769.898}{\nano\meter}.  The original BiSON
spectrophotometer is described in detail by~\citet{brookesphd}
and~\citet{1978MNRAS.185....1B}.  The design philosophy has changed
little over the years, with differences only in the type of chassis
and style of magnet/oven assembly.

Figure~\ref{fig:fibre_rss} shows an optical schematic of the new
miniaturised fibre-fed spectrophotometer.  Over the next few
subsections we describe an end-to-end system model, from the objective
optics, through optical signal processing, to the final detectors.

%%%%%%%%%%%%%%%%%%%%%%%%%%%%%%%%%%%%%%%%%%%%%%%%%%%%%%%%%%%%%%%%%%%%%%%%%%%%%%%%

\subsection{Optical Power Budget}

We start by considering the power available at the entrance to the
instrument.  The solar constant is~\SI{1365}{\watt\per\meter\squared},
measured at the top of Earth's atmosphere, and the contribution to
this near our target wavelength of~\SI{770}{\nano\meter}
is~\SI{1.22}{\watt\per\meter\squared\per\nano\meter}.  To determine
the power reaching the ground we must take into account atmospheric
extinction.

At~\SI{770}{\nano\meter} atmospheric extinction is dominated by the
contribution of aerosols.  The aerosol optical depth at the BiSON site
at Las~Campanas, Chile, is typically \num{0.03}~magnitudes per
airmass, whereas the BiSON site at Iza\~na, Tenerife, can reach as
high as \num{0.5}~magnitudes per airmass during periods of high
Saharan dust~\citep{2017AJ....154...89H}.  At one airmass with an
aerosol optical depth of~\num{0.03}, the power spectral density at sea
level near~\SI{770}{\nano\meter} is
\SI{1.18}{\watt\per\meter\squared\per\nano\meter}.  If we estimate six
airmasses with an aerosol optical depth of~\num{0.1}, the typical
maximum value outside of periods of high Saharan dust, this reduces
the power spectral density at sea level to
\SI{0.67}{\watt\per\meter\squared\per\nano\meter}.  When modelling
system response, we assume an optical power spectral density budget of
\SI{0.9}{\watt\per\meter\squared\per\nano\meter} as a nominal
mid-range value.

%%%%%%%%%%%%%%%%%%%%%%%%%%%%%%%%%%%%%%%%%%%%%%%%%%%%%%%%%%%%%%%%%%%%%%%%%%%%%%%%

\subsection{Objective Optics}

\begin{figure}
  \includegraphics[width=\columnwidth]{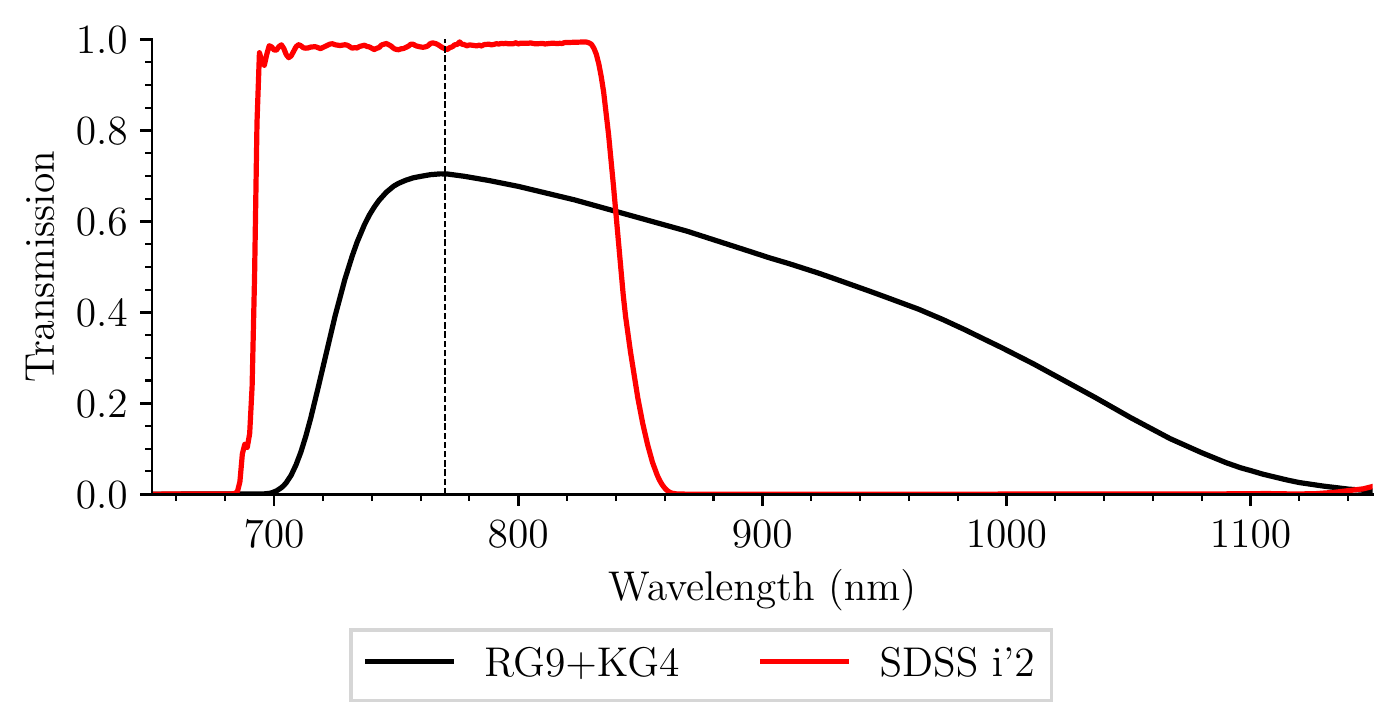}
  \caption{Profiles of the SDSS {i'2} filter, and the original RG9+KG4
    filters.  The dashed vertical line indicates
    the~\SI{770}{\nano\meter} operational wavelength.}
\label{fig:red_filter}
\end{figure}

A deep-red filter is the first element in the system, and is used to
relieve much of the infra-red thermal load on the instrument.
Historically, this has been a combination of a Schott~RG9 at
\SI{3}{\milli\meter}~thickness, and a Schott~KG4 at
\SI{1}{\milli\meter}~thickness, giving a combined transmission
of~\SI{70}{\percent} at~\SI{770}{\nano\meter}.  Schott~KG4 has been
discontinued and is obsolete.  The most common filter set in use today
is the photometric system designed by~\citet{1996AJ....111.1748F} for
the Sloan Digital Sky Survey ({SDSS}).  The SDSS {i'2} filter is an
ideal replacement, with a bandwidth of~\SIrange{695}{844}{\nano\meter}
providing better infra-red rejection, and transmission
at~\SI{770}{\nano\meter} of~\SI{98}{\percent}.
Figure~\ref{fig:red_filter} shows the profile of the SDSS {i'2}
filter, and the combined RG9+KG4 profile.

Objective lens~L1 has~\SI{25.4}{\milli\meter} diameter
and~\SI{30}{\milli\meter} focal length.  The solar field angle at
perihelion is~\SI{0.542}{\degree}, and this is the minimum acceptance
angle required to view the whole extent of the solar disc.  Lens~L1
projects a~\SI{0.6}{\milli\meter}~image of the solar disc onto the end
of an optical fibre with \SI{1.0}{\milli\meter}~core diameter.  The
image is larger than~\SI{0.3}{\milli\meter} predicted by first-order
optics due to the fast focal ratio and imperfect correction of
aberrations, but well within the acceptance of the fibre.  The lens
clear aperture is~\SI{22.8}{\milli\meter}
collecting~\SI{0.37}{\milli\watt\per\nano\meter} based on our optical
power budget.

The combined collection optics are small and light enough, and have
sufficient tolerance to guiding errors, to be used with any
inexpensive amateur telescope mount~\citep{halephd,10.1117/12.2561282}.

%%%%%%%%%%%%%%%%%%%%%%%%%%%%%%%%%%%%%%%%%%%%%%%%%%%%%%%%%%%%%%%%%%%%%%%%%%%%%%%%

\subsection{Fibre Coupling}

A off-the-shelf \SI{1000}{\micro\meter}~multimode fibre is used to
couple the collection optics to the instrument.  The numerical
aperture is~\num{0.39} producing an acceptance angle
of~\SI{23}{\degree}, matching the cone of light produced by the
objective lens.  The large core diameter was selected in order to both
improve the light throughput, and to prevent the focal ratio of the
collection optics becoming unfeasibly small.

The optical fibre scrambles the image of the Sun that is focused on
the input of the fibre by the collection optics.  Since BiSON observes
the Sun-as-a-star this is a desirable addition to the design, as it
reduces sensitivity to solar rotation, instrumental vignetting, and
almost completely eliminates sensitivity to guiding error.  Since an
image of the Sun is focused onto the end of the fibre it is
essential, as with all optics in a focal plane, that the fibre end
remains completely clean and undamaged.  A piece of dust or dirt on
the fibre end would cause variable transmission across the input, and
then motion of the solar image due to guiding errors would cause
different parts of the solar image to be preferentially transmitted.
The objective lens and fibre connection are sealed inside an optical
tube assembly.

The fibre output is collimated by lens~L2, with the same properties as
the objective lens~L1.  The collimated beam through the instrument has
a diameter of~\SI{22.8}{\milli\meter} and divergence
of~\SI{1}{\degree}.  The power entering the spectrophotometer is
estimated at~\SI{0.34}{\milli\watt\per\nano\meter}.  Due to the beam
divergence there is a some vignetting through the instrument, and this
is minimised by ensuring all components are as close together as
possible.  Since the beam is uniform after passing through the optical
fibre, the small amount of vignetting is simply a loss of power rather
than a loss of information.

%%%%%%%%%%%%%%%%%%%%%%%%%%%%%%%%%%%%%%%%%%%%%%%%%%%%%%%%%%%%%%%%%%%%%%%%%%%%%%%%

\subsection{Optical Signal Processing}

The first stage of optical signal processing is an interference filter
with~\SI{1.5}{\nano\meter} bandwidth.  This is used to isolate the
potassium~D1 absorption line at~\SI{770}{\nano\meter} from the nearby
D2 absorption line at~\SI{766}{\nano\meter}, which is heavily
contaminated by atmospheric effects.

The potassium vapour cell is placed in a longitudinal magnetic field
which causes the line to be Zeeman-split into two components, where
the separation is dependent on the magnetic field strength.  The two
components of the Zeeman split line interact only with circularly
polarised light, and since the two components have opposite circular
polarisation it allows the LCD rotator and fixed quarter-wave plate to
select which component should be observed.  Splitting the lab
reference-frame in this way produces two working-points of
measurement, and so allows discrimination between red- and blue-shift.
It also moves the measurements onto the the steeper parts of the wings
of the solar absorption line, improving Doppler velocity sensitivity
and allows measurement even during periods crossing zero line-of-sight
velocity.

Lens {L3} forms an image of the fibre at the centre of the vapour
cell.  The requirements are that the focal length is long enough to
ensure the beam has converged sufficiently before entering the
aperture in the side of the magnet, but short enough so as not to
cause unnecessarily high magnification of the image.  A large image
and high divergence would increase the likelihood of direct scattering
from the glass cell walls, which would easily obscure the signal from
atomic scattering.  A focal length of~\SI{100}{\milli\meter} is used,
producing a~\SI{3.3}{\milli\meter} diameter image of
the~\SI{1}{\milli\meter} fibre at the centre of the cell.  The power
entering the cell is estimated at~\SI{0.08}{\milli\watt}, and is now
effectively monochromatic.

The data product is an intensity-normalised ratio $R$ formed from the
two instrumental passbands of light scattered from the vapour cell,
\begin{equation}
  R = \frac{I_{\mathrm{b}} - I_{\mathrm{r}}}{I_{\mathrm{b}} + I_{\mathrm{r}}} ~,
\end{equation}
where $I_{\mathrm{b}}$ and $I_{\mathrm{r}}$ are the intensities
measured at the blue and red wings of the solar absorption line,
respectively.  This ratio is closely related to the Earth-Sun
line-of-sight velocity, and can be calibrated according
to~\citet{1995A&AS..113..379E} and~~\citet{doi:10.1093/mnras/stu803}.

%%%%%%%%%%%%%%%%%%%%%%%%%%%%%%%%%%%%%%%%%%%%%%%%%%%%%%%%%%%%%%%%%%%%%%%%%%%%%%%%

\subsection{Vapour Cell and Magnetic Field Strength}

\begin{figure}
  \includegraphics[width=\columnwidth]{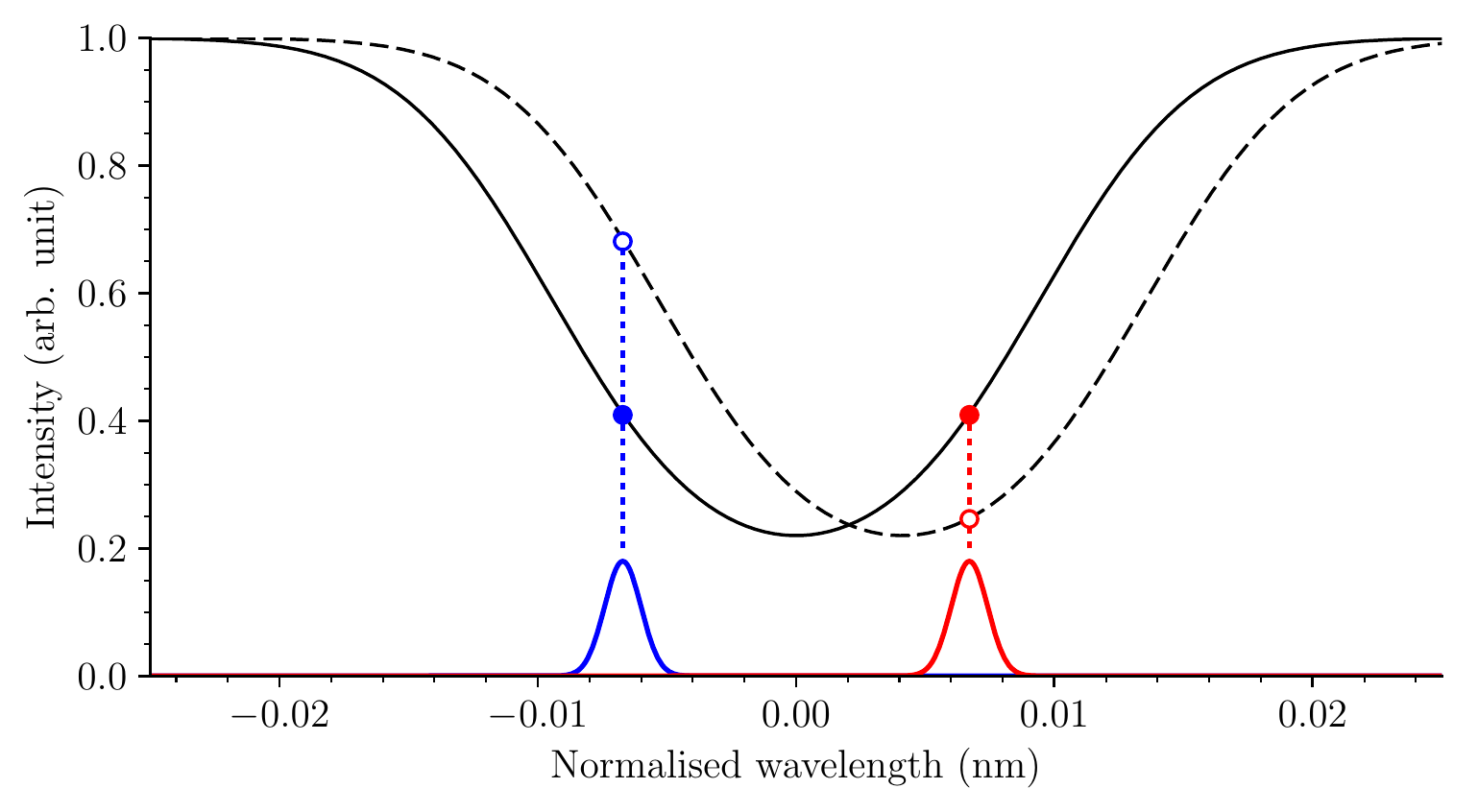}
  \caption{The potassium solar absorption line is shown with both zero
    line-of-sight velocity (solid line), and also red-shifted
    by~\SI{1600}{\metre\per\second} (dashed line) to demonstrate the
    effect on the scattering intensity.  The two absorption components
    are split by a~\SI{0.18}{\tesla} longitudinal magnetic field
    producing a separation of approximately~\SI{0.013}{\nano\metre},
    or~\SI{5.2}{\kilo\metre\per\second} in terms of Doppler velocity.}
  \label{fig:splitting}
\end{figure}

\begin{figure}
  \includegraphics[width=\columnwidth]{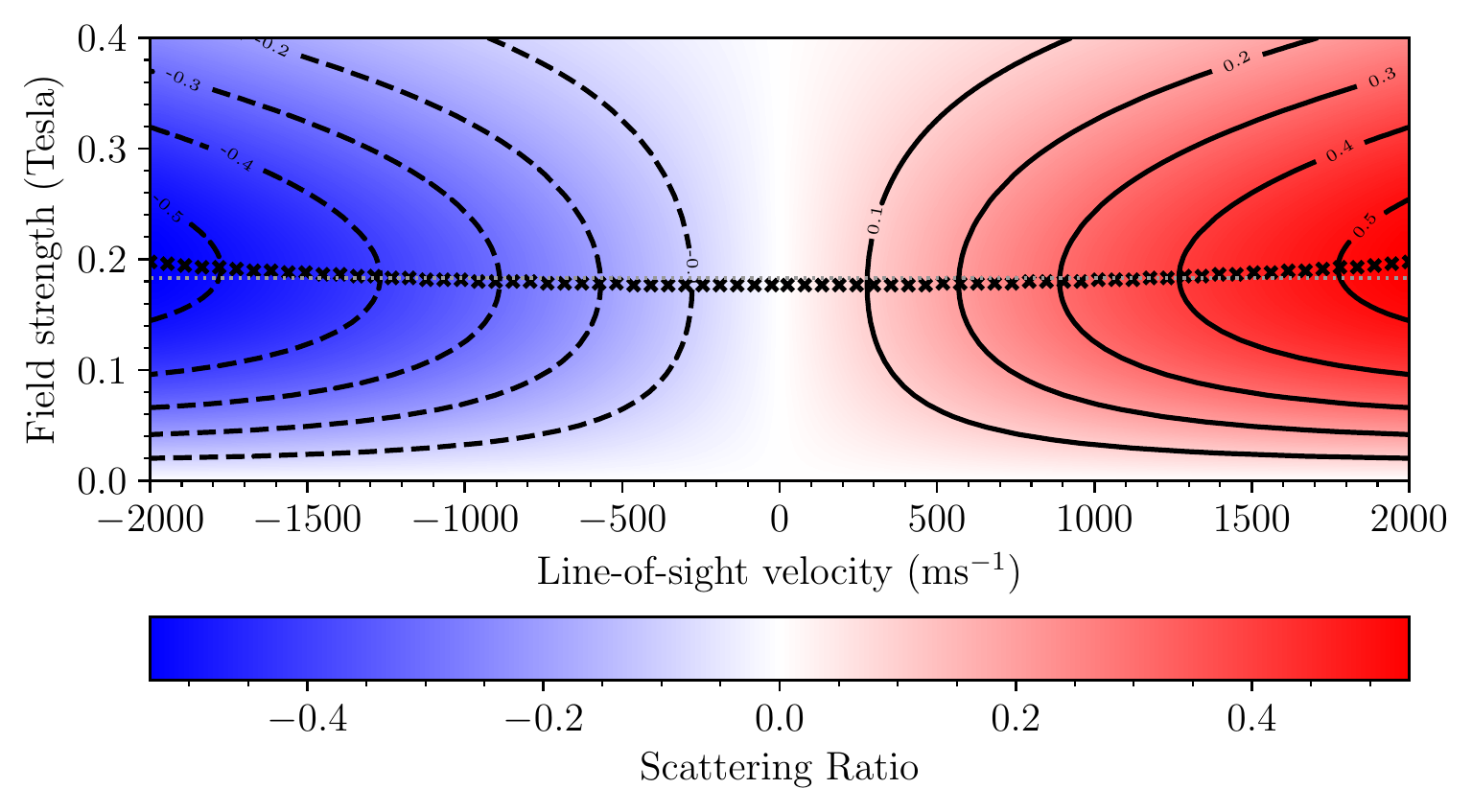}
  \caption{Magnetic field strength optimisation.  The crosses indicate
    the point of maximum ratio for each velocity and therefore the
    ideal magnetic field strength.  The faint horizontal grey dotted
    line indicates the mean ideal field strength of~\SI{0.18}{\tesla}.}
  \label{fig:magcal}
\end{figure}

The potassium~D1 absorption line profile is shown in
Figure~\ref{fig:splitting}, modelled according
to~\citet{underhillmsc}.  The solar line-of-sight velocity shifts
between approximately~\SIrange{-300}{1600}{\metre\per\second} over one
year, and this gives lower and upper bounds on the acceptable magnetic
field strength.

The lower limit to the applied magnetic field is set by the
requirement that the red and blue components are both fully separated,
and always on opposite sides of the line at all line-of-sight
velocities.  The upper limit is set by the blue component moving
completely off the absorption line and into the continuum at maximum
offset.  The potential range of magnetic field strength is therefore
between~\SIrange{0.1}{0.3}{\tesla}.

Figure~\ref{fig:magcal} shows the predicted scattering ratio over a
range of velocity offsets and magnetic field strengths, drawn
symmetrically for completeness.  The ideal field strength maximising
velocity sensitivity is~\SI{0.18}{\tesla}, which splits the two
components by~\SI{0.013}{\nano\metre},
or~\SI{5.2}{\kilo\metre\per\second} in terms of Doppler velocity.

\begin{figure}
    \subfloat[BiSON magnet housing]{\includegraphics[width=0.5\columnwidth]{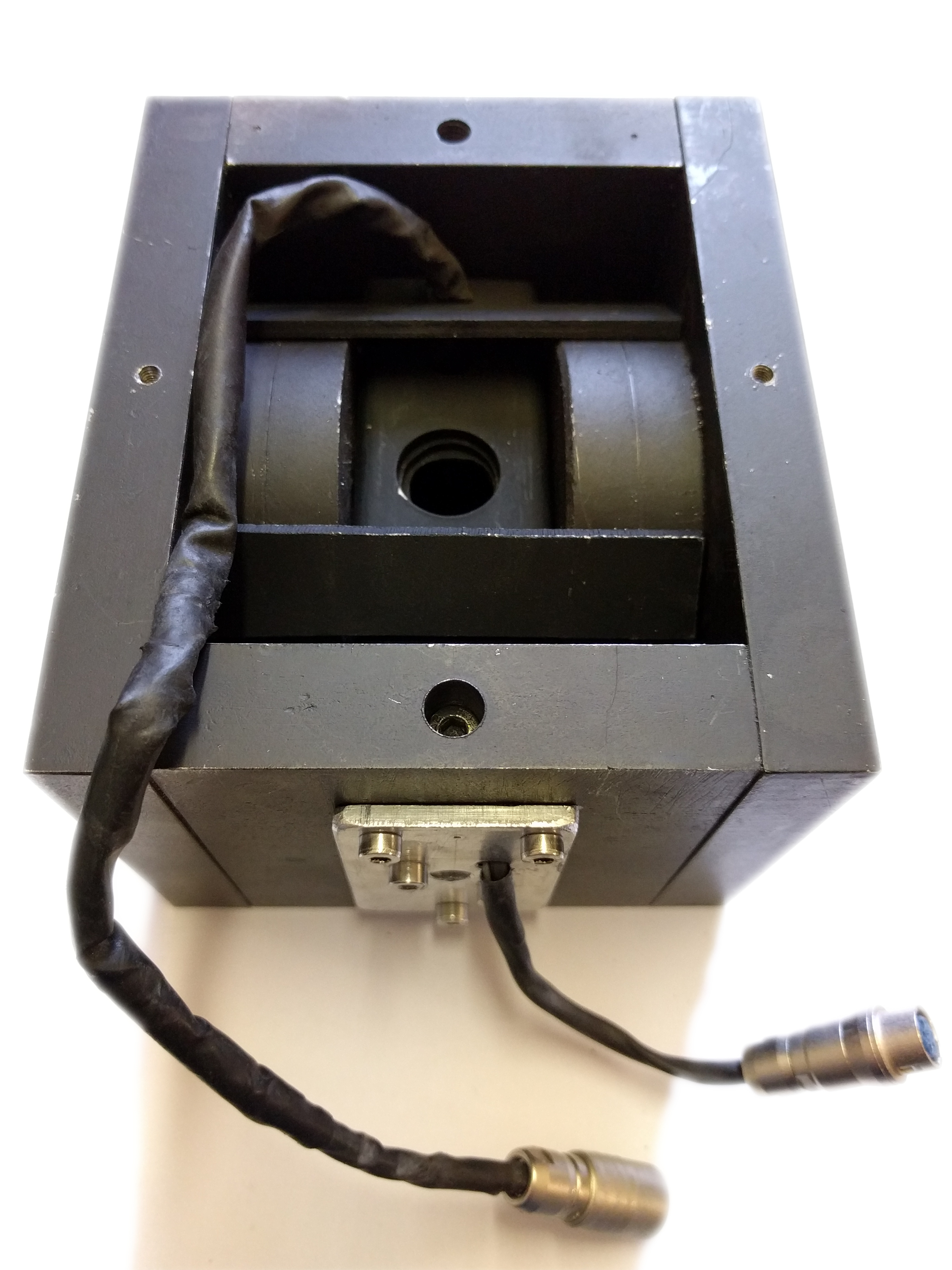}
                                    \label{subfig:bison-magnet}}%
    \quad
    \subfloat[BiSON vapour cell oven]{\includegraphics[width=0.5\columnwidth]{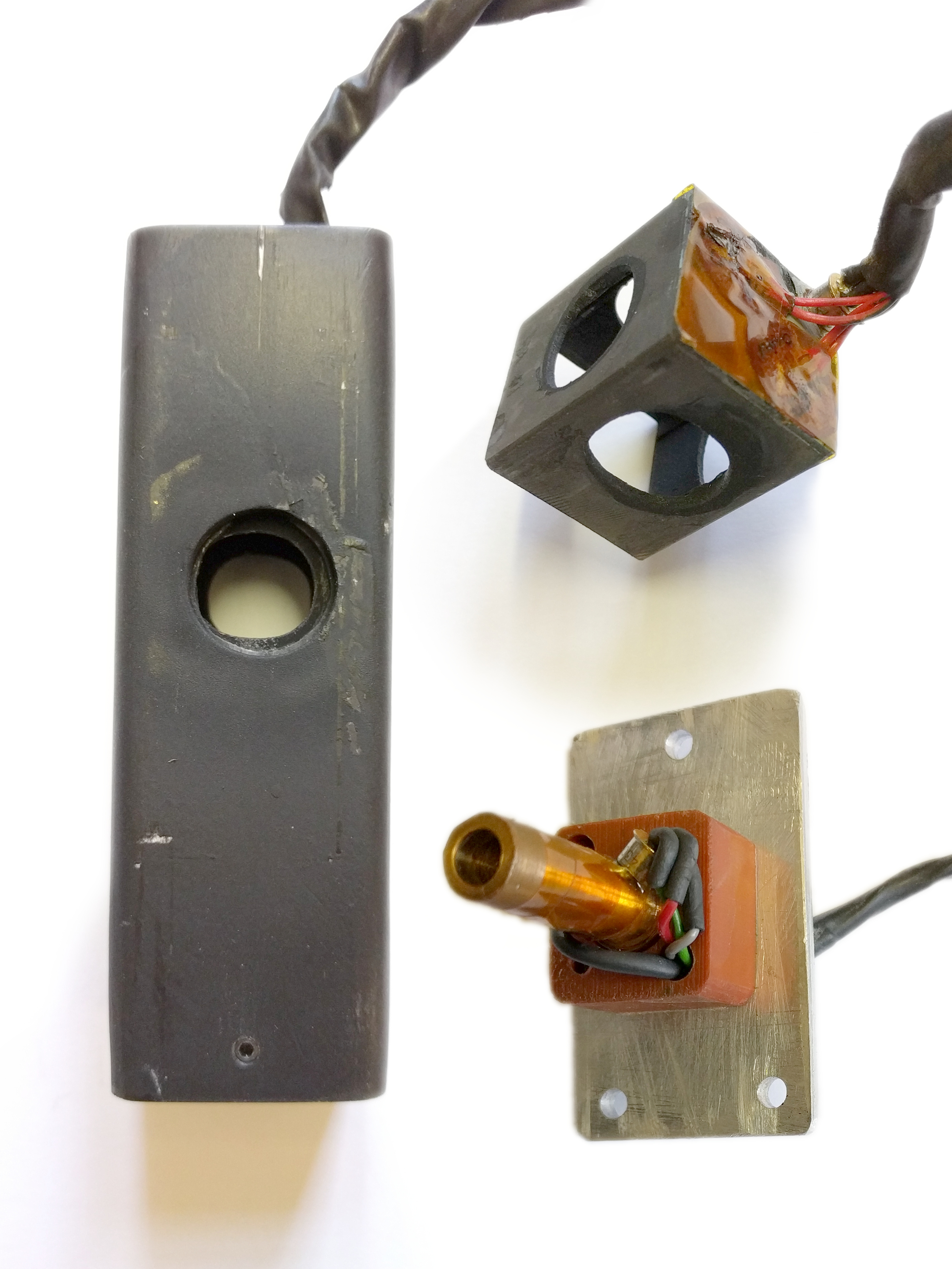}
                                     \label{subfig:bison-oven}}%
    \caption{An example BiSON magnet and vapour cell
      oven. \protect\subref{subfig:bison-magnet} shows the magnet and
      yoke, with the cell oven installed inside.
      \protect\subref{subfig:bison-oven} shows the cell oven,
      including the separate cube and stem heaters.}
    \label{fig:bison_magnet_cell}%
\end{figure}

Figure~\ref{fig:bison_magnet_cell} shows the magnet and cell oven
assembly.  It is possible to estimate the power reaching the
scattering detectors by considering the geometry of the cell, the
scattering cross section, and the collection optics.  The scattering
intensity is maximised at a cell stem temperature of
approximately~\SIrange{80}{90}{\celsius}, at which temperature the
optical depth is approximately unity when measured from the front of
the cell to the centre~\citep{2020JPhB...53h5003H}.  We first scale
by,
\begin{equation}
 I_1 = I_0 \, e^{-1} ~,
\end{equation}
making an assumption of single-scattering.  This is acceptable since
we require only a simple power estimate. We then scale by the
scattering cross-section,
\begin{equation}
 I_2 = I_1 \, \frac{\lambda_{\mathrm{cs}}}{\lambda_{\mathrm{beam}}} ~,
\end{equation}
where $\lambda_{\mathrm{cs}}$ is the scattering cross-section of
approximately~\SI{4}{\pico\meter} and $\lambda_{\mathrm{beam}}$ is the
incoming beam bandwidth of approximately~\SI{1}{\nano\meter}.  The
cell scattering is isotropic, and so if we assume a spherical cell we
can make use of the inverse square law to calculate the power incident
at the collection optics,
\begin{equation}
 I_3 = I_2 \, \frac{d^2}{D^2} ~,
\end{equation}
where $d$ is the image diameter at the centre of the cell of
approximately~\SI{3.3}{\milli\meter} and $D$ is the distance to the
collection optics of approximately~\SI{9}{\milli\meter}.  Finally we
need to scale by the aperture area,
\begin{equation}
 I_{\mathrm{det}} = I_3 \, \frac{A_{\mathrm{det}}^2}{A_{\mathrm{beam}}^2} ~,
\end{equation}
where $A_{\mathrm{beam}}$ is the beam aperture of
approximately~\SI{20}{\milli\meter} and $A_{\mathrm{det}}$ is the
detector aperture of~\SI{6}{\milli\meter}.  The final intensity at the
detector, $I_{\mathrm{det}}$, is approximately~\SI{1.5}{\nano\watt} if
measuring in the solar continuum.  The solar potassium absorption line
has an absorption depth of around~\SI{77}{\percent}, and so the final
value is between~\SI{0.35}{\nano\watt} and~\SI{1.4}{\nano\watt}
depending on the line-of-sight velocity offset.  We assume a
nominal~\SI{1}{\nano\watt} of scattering power.

%%%%%%%%%%%%%%%%%%%%%%%%%%%%%%%%%%%%%%%%%%%%%%%%%%%%%%%%%%%%%%%%%%%%%%%%%%%%%%%%

\subsection{Transmission Monitor}

It is useful to monitor the light that is not scattered by the cell
and is transmitted directly through the spectrophotometer.  This
measurement allows the quality of beam polarisation to be monitored,
and also to estimate the daily atmospheric extinction
coefficient~\citep{2017AJ....154...89H}.

Lens {L4} is a~\SI{50}{\milli\meter} focal length bi-convex lens
positioned to focus light exiting from the cell into an image on the
transmission detector, via a fixed linear polariser and quarter-wave
plate.  We determined that the optical power entering the vapour cell
is~\SI{0.08}{\milli\watt}, and this can be also be considered to be
the total light exiting the cell since very little power is removed
from the beam by resonance scattering.  The power at the transmission
detector is~\SI{0.07}{\milli\watt} when the input and output
polarisation of the whole system is uncrossed, and essentially zero
when crossed.  The filters and the photodiode are mounted with a small
wedge angle in order to prevent back-reflections into the cell.

%%%%%%%%%%%%%%%%%%%%%%%%%%%%%%%%%%%%%%%%%%%%%%%%%%%%%%%%%%%%%%%%%%%%%%%%%%%%%%%%

% -*- coding: utf-8 -*-
%
% NOISE.TEX
%
%   Steven Hale
%   2022 February 11
%   Birmingham, UK
%
% Noise Sources
%

%%%%%%%%%%%%%%%%%%%%%%%%%%%%%%%%%%%%%%%%%%%%%%%%%%%%%%%%%%%%%%%%%%%%%%%%%%%%%%%%

\section{Noise Sources}
\label{s:noise}

The noise characteristics of RSS instruments have been investigated by
several authors in terms of time-dependent effects on the science
extracted from the observations --
see,~e.g.,~\citet{1976A&A....50..221G, 1978MNRAS.185....1B,
  1989A&A...222..361A, 1989ApJ...345.1088H, 1991SoPh..133...43H,
  2005MNRAS.359..607C}.  Here, we consider the contributions to noise
on the whole envelope.

\subsection{Atmospheric Scintillation}

The effect of atmospheric scintillation on the measured intensity has
a maximum of
around~\SI{15}{\nano\watt\per\sqrthz}~\citep{2020PASP..132c4501H}, and
this is the dominant noise source.  The~\SI{-3}{\decibel} point
(noise power reduced to~\SI{50}{\percent}) typically occurs at
around~\SI{5}{\hertz}, and the~\SI{-6}{\decibel} point (noise power
reduced to~\SI{25}{\percent}) typically occurs at
between~\SIrange{5}{15}{\hertz}.  The~\SI{-10}{\decibel} point (noise
power reduced to~\SI{10}{\percent}) typically occurs
above~\SI{20}{\hertz}~\citep{2020PASP..132c4501H}.

BiSON instrumentation reduces the effect of scintillation noise by
forming a velocity ratio from the measurements of the two working
points on the solar absorption line.  The level of noise reduction is
determined by how quickly the system can switch between polarisation
states -- faster switching requires shorter exposure times, with
observation of each wing closer to being simultaneous, and the
instrument ``sees'' less atmospheric turbulence.

The original Mark-I instrument switches at~\SI{0.5}{\hertz} and little
of the atmospheric scintillation is eliminated.  Newer BiSON
instrumentation uses a bespoke Pockels-effect cell and drive
electronics to achieve a switching rate of~\SI{90.9}{\hertz}, with
\SI{15}{\percent}~dead time due to stabilisation delays and data
readout within each integration period.  At this switching rate the
atmospheric scintillation power has decayed below~\SI{10}{\percent}
and is less than~\SI{5}{\nano\watt\per\sqrthz}.

We saw earlier that the new miniature instrument uses an off-the-shelf
LCD to control polarisation state.  The LCD rotator requires up
to~\SI{27}{\milli\second} to change state~\citep[submitted]{detectorbandwidth},
and so a switching rate of around~\SI{5}{\hertz} can be obtained with
dead time of~\SI{20}{\percent}.  This can be increased to
around~\SI{10}{\hertz} with~\SI{50}{\percent}~dead time.  Due to the
slower switching rate we can expect the white noise level due to
atmospheric scintillation to rise to
between~\SIrange{7.5}{10}{\nano\watt\per\sqrthz}, depending on natural
daily variability.

%%%%%%%%%%%%%%%%%%%%%%%%%%%%%%%%%%%%%%%%%%%%%%%%%%%%%%%%%%%%%%%%%%%%%%%%%%%%%%%%

\subsection{Guiding Noise}

In traditional BiSON instrumentation where the Sun is imaged inside
the vapour cell, an additional source of noise is caused by guider
errors.  Guider fluctuations cause the image of the Sun to move around
inside the vapour cell, which changes the vapour optical depth seen by
the scattering detectors, and this in turn causes changes in the
measured velocity ratio.  It is difficult to determine a quantitative
measure of this effect.

As discussed earlier, the new instrument focuses an image of the Sun
onto the end of an optical fibre.  The output of the fibre is uniform
regardless of the absolute position of the image provided that the
image falls entirely within the core of the fibre.  The constraint on
guiding accuracy is relaxed and guider noise is considered to be
eliminated completely, so long as the error is not so large as to
cause vignetting.

%%%%%%%%%%%%%%%%%%%%%%%%%%%%%%%%%%%%%%%%%%%%%%%%%%%%%%%%%%%%%%%%%%%%%%%%%%%%%%%%

\subsection{Electronic Noise}

\begin{figure}
  \includegraphics[width=\columnwidth]{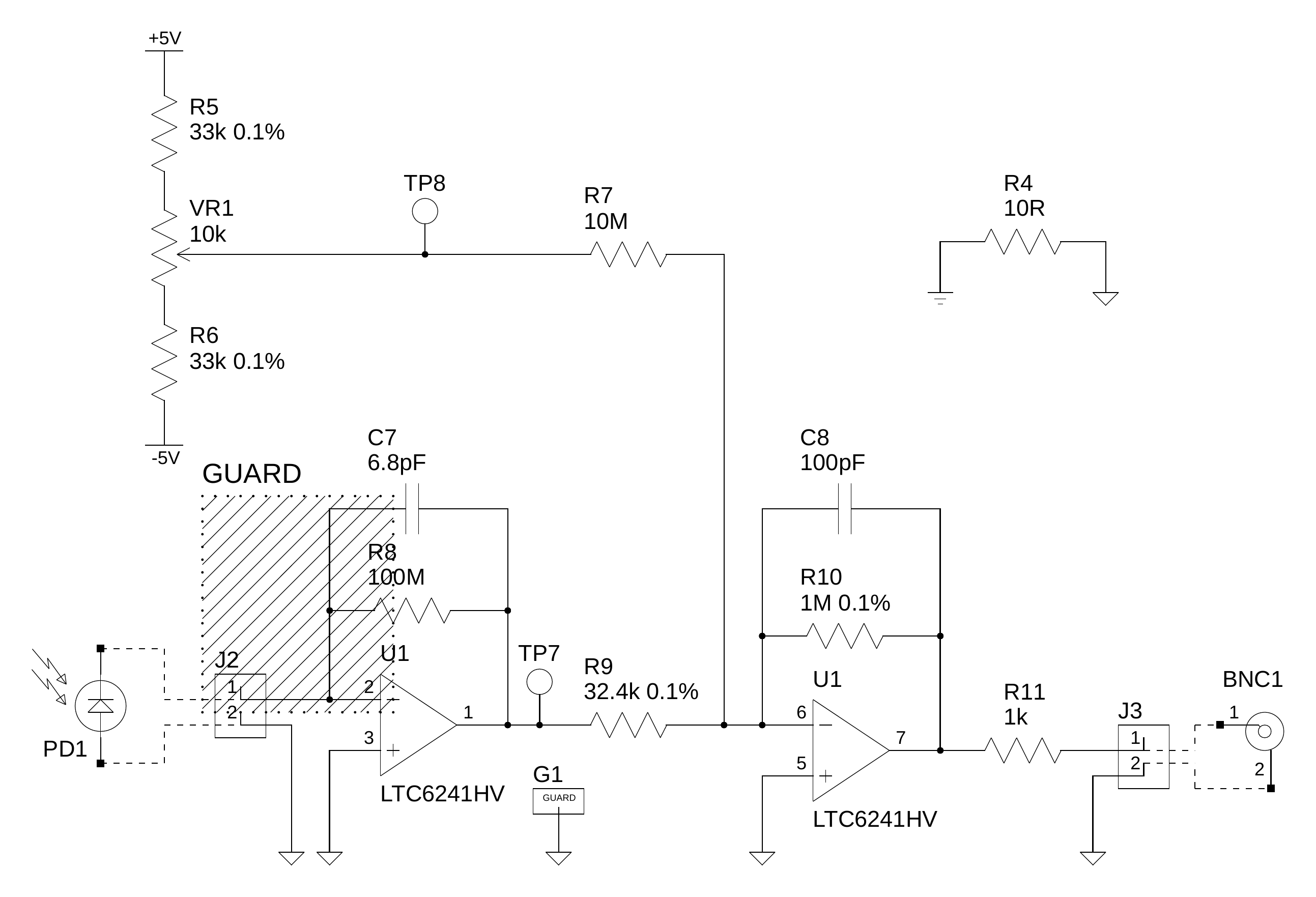}
  \caption{Scattering detector transimpedance amplifier schematic.}
  \label{fig:amplifier}
\end{figure}

There are two scattering detectors, one on each side of the cell
perpendicular to the beam.  We saw earlier that the expected optical
intensity is between~\SI{0.35}{\nano\watt} and~\SI{1.4}{\nano\watt}.
The detectors use a photodiode with responsivity
at~\SI{770}{\nano\meter} of~\SI{0.45}{\ampere\per\watt}.  The expected
photocurrent is of the order of~\SIrange{160}{630}{\pico\ampere},
which requires high amplification to raise the signal to a suitable
level for the input to the digitizers.  In comparison to the five
minute solar oscillations, relatively high bandwidth of a few hundred
Hz is also required due to multiplexing multiple polarisation states
creating transient signal state changes.  The amplifier schematic
itself is relatively simple, shown in Figure~\ref{fig:amplifier}.
However, with such high impedance circuits great care must be taken in
the design and construction to ensure the expected performance levels
are achieved.

An amplifier with low noise characteristics was selected for both
amplification stages.  The input bias current typical value is
just~\SI{0.2}{\pico\ampere}, the input noise
is~\SI{0.56}{\femto\ampere\per\sqrthz}, and it has a low temperature
coefficient at~\SI{2.5}{\micro\volt\per\celsius}.  It is a bipolar
rail-to-rail device allowing the output stage to swing
within~\SI{30}{\milli\volt} of either supply rail and so maximise the
signal dynamic range.  A total of three different grounds are defined
in the circuit to isolate the signal from noise sources as much as
possible.  The sensitive high-impedance connection between the
photodiode output and the inverting input of the transimpedance
amplifier is protected by a low-impedance signal guard ring.  The
soldermask is removed from this section of the board to avoid creating
a leakage bridge over the guard ring.  This is essential to limit
noise, since even a~\SI{10}{\giga\ohm} resistance across the PCB
between a~\SI{5}{\volt} supply trace and a high-impedance input would
add~\SI{500}{\pico\ampere} of leakage current -- equal to the signal
we are are trying to amplify and three orders of magnitude higher than
the input bias current of the operational amplifier.

\begin{figure}
  \includegraphics[width=\columnwidth]{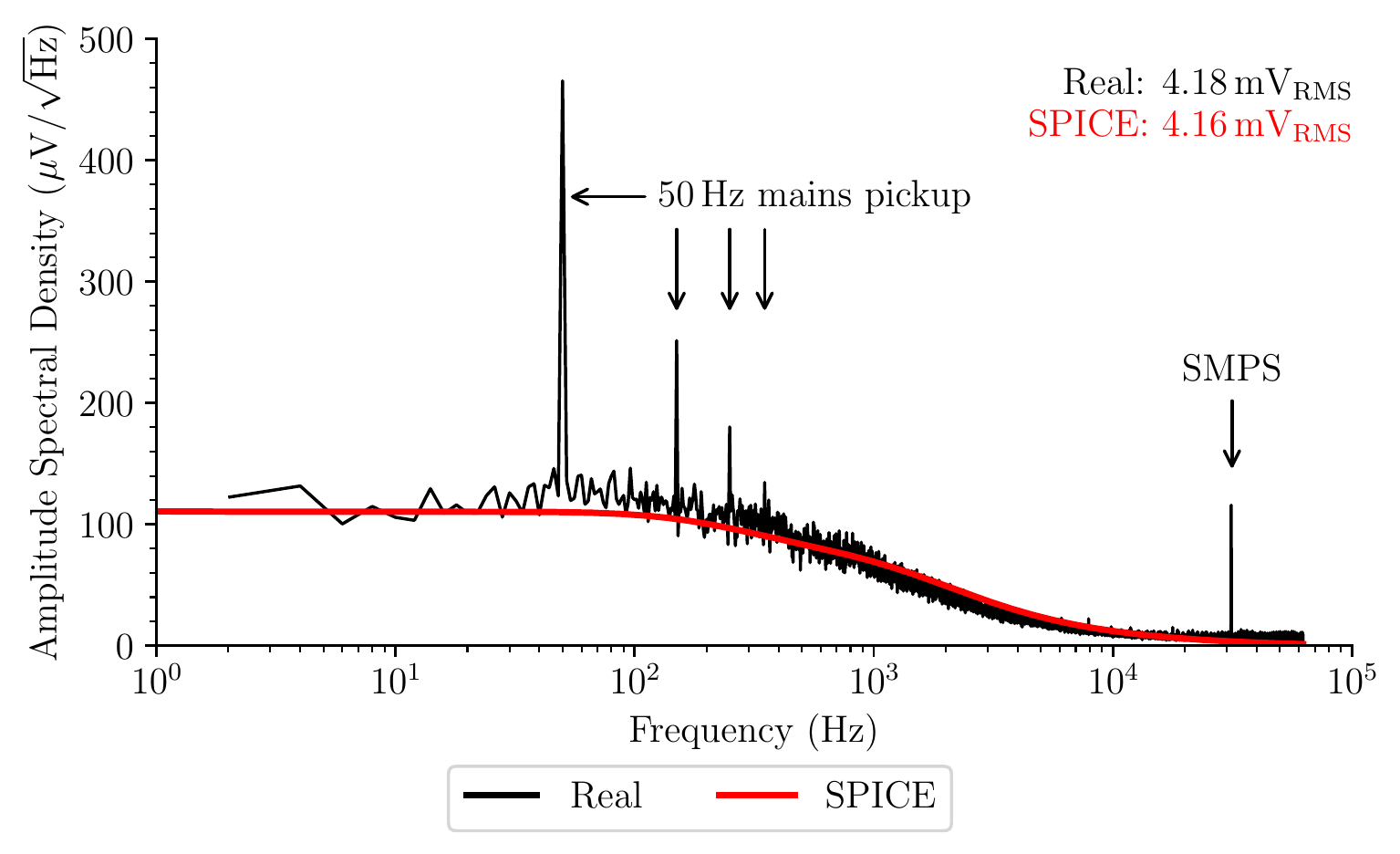}
  \caption{Measured scattering detector noise, along with the expected
    noise profile from SPICE simulation.  The physical detector
    suffers from some minor mains pickup at~\SI{50}{\hertz}, and a
    peak around~\SI{30}{\kilo\hertz} due to powering the detector from
    a switched-mode power-supply (SMPS), but broadly matches
    expectation.}
  \label{fig:scattering_isolated}
\end{figure}

The noise performance of the detector was simulated using SPICE, the
industry standard electrical simulation tool~\citep{Nagel:M382,Nagel:M520}.
Figure~\ref{fig:scattering_isolated} shows the simulated detector in
red, and data from the real detector in black.  The real detector
suffers from some slight mains pickup at~\SI{50}{\hertz}, and the peak
around~\SI{30}{\kilo\hertz} is due to powering the detector from a
switched-mode power-supply.  The noise level is minor and the
performance matches expectation.  The RMS noise amplitude is
approximately~\SI{4.2}{\milli\vrms}.  The typical noise value
of~\SI{100}{\micro\volt\per\sqrthz} can be rescaled by the amplifier
gain enabling the noise equivalent power (NEP) to be calculated.  This
is the equivalent signal that would be required to produce the same
output from an ideal noiseless detector.  The electronic NEP
is~\SI{0.07}{\pico\watt\per\sqrthz}, several orders of magnitude less
than that due to atmospheric scintillation.

The bandwidth~\SI{-3}{\decibel} point (noise power reduced
to~\SI{50}{\percent}) occurs at~\SI{239}{\hertz}, producing a slew
duration of about~\SI{4}{\milli\second} which is sufficient to pass
the transient changes that occur during LCD polarisation state
changes.

%%%%%%%%%%%%%%%%%%%%%%%%%%%%%%%%%%%%%%%%%%%%%%%%%%%%%%%%%%%%%%%%%%%%%%%%%%%%%%%%

\subsection{Photon Shot Noise}

The typical incident flux on the scattering detectors is of the order
of~\SI{1}{\nano\watt}.  For~\SI{1}{\nano\watt} of light
at~\SI{770}{\nano\meter} this is equivalent to approximately
\num{4e9}~photons per second.  The standard deviation of shot noise is
equal to the square-root of the average number of events, and
converting this back to power gives a value
of~\SI{0.02}{\pico\watt_{RMS}}.  At the detector output this
is~\SI{0.03}{\milli\vrms}, two orders of magnitude smaller than the
electronic noise.

%%%%%%%%%%%%%%%%%%%%%%%%%%%%%%%%%%%%%%%%%%%%%%%%%%%%%%%%%%%%%%%%%%%%%%%%%%%%%%%%

\subsection{Thermal Noise}

The performance of various stages of the the instrument are sensitive
to their operating temperature.  We actively control the working
temperature of the the potassium cell stem and cube, the LCD rotator,
and the two scattering photodiodes.  The narrow-band interference
filter is thermally stabilised on traditional BiSON instrumentation,
in order to limit intensity changes resulting from movement of the
passband and spectral fringes.  When the instrument is operated in a
thermally controlled environment and separated from the collection
optics in direct sunlight, precise thermal control of filters is not
required.

The vapour cell has the most significant thermal contribution to the
noise.  Variations in the vapour cell stem temperature introduce
changes in the vapour temperature, and this affects the scattering
intensity due to changes in the vapour optical depth.

The temperature control system can achieve a mean stem temperature
of~\SI{90}{\celsius} with a standard deviation
of~\SI{0.7e-3}{\celsius}.  Using the vapour cell model
from~\citet{2020JPhB...53h5003H} this can be converted into scattering
intensity noise.  Assuming a detector aperture of~\SI{6}{\milli\meter}
and scaling for a typical~\SI{1}{\nano\watt} intensity, the effect of
thermal noise on the vapour cell is~\SI{0.2}{\pico\watt\per\sqrthz},
which is of a similar order to the electronic noise.

%%%%%%%%%%%%%%%%%%%%%%%%%%%%%%%%%%%%%%%%%%%%%%%%%%%%%%%%%%%%%%%%%%%%%%%%%%%%%%%%

% -*- coding: utf-8 -*-
%
% ACQUISITION.TEX
%
%   Steven Hale
%   2022 February 11
%   Birmingham, UK
%
% Data Acquisition
%

%%%%%%%%%%%%%%%%%%%%%%%%%%%%%%%%%%%%%%%%%%%%%%%%%%%%%%%%%%%%%%%%%%%%%%%%%%%%%%%%

\section{Data Acquisition}
\label{s:acquisition}

The data acquisition system is based around a number 24-bit
delta-sigma ($\Delta\Sigma$) analogue to digital converters.  The ADCs
are an integrated package containing a programmable gain amplifier
(PGA), a second-order $\Delta\Sigma$ modulator, a programmable digital
low-pass filter, and a micro-controller with several control
registers.

The $\Delta\Sigma$ modulator over-samples at high rate, and the rate is
then reduced by the internal digital low-pass filter.  This means that
the digital quantisation noise is pushed up to higher frequencies and
then removed by the low-pass decimation filter producing an effect
known as noise shaping.  The ADC can be considered to be effectively
noiseless.

Data are read-out by a Raspberry~Pi single-board computer over the
on-board Serial Peripheral Interface (SPI) bus.  The Raspberry~Pi is a
small single-board computer based around a Broadcom System-on-a-Chip
(SoC) with an integrated ARM compatible CPU, and runs a version of the
GNU/Linux operating system.  An SoC paired with the integrated PGA,
modulator, and low-pass filter, allows the whole acquisition system to
be small and self-contained.

The calibration of Doppler shift to velocity is discussed
by~\citet{1995A&AS..113..379E}, with an improved data pipeline that
includes correction for atmospheric differential extinction described
by~\citet{doi:10.1093/mnras/stu803}.  The subsequent concatenation of
data from multiple sites is discussed by~\citet{1997A&AS..125..195C}
and \citet{halemphil,halephd}.

%%%%%%%%%%%%%%%%%%%%%%%%%%%%%%%%%%%%%%%%%%%%%%%%%%%%%%%%%%%%%%%%%%%%%%%%%%%%%%%%

% -*- coding: utf-8 -*-
%
% COMMISSIONING.TEX
%
%   Steven Hale
%   2022 February 11
%   Birmingham, UK
%
% Prototype Commissioning
%

%%%%%%%%%%%%%%%%%%%%%%%%%%%%%%%%%%%%%%%%%%%%%%%%%%%%%%%%%%%%%%%%%%%%%%%%%%%%%%%%

\section{Prototype Commissioning}
\label{s:commissioning}

\begin{figure}
  \includegraphics[width=\columnwidth]{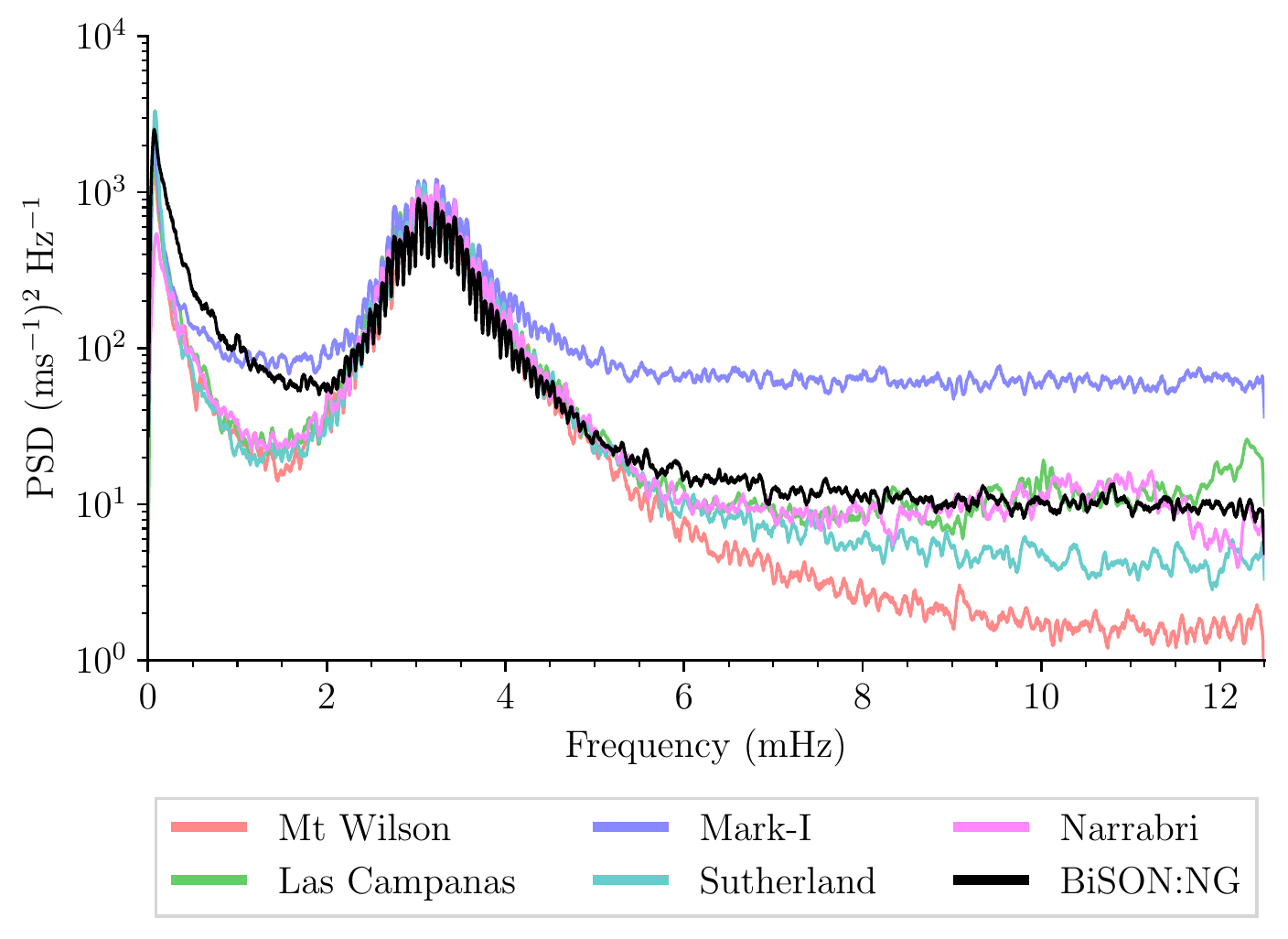}
  \caption{Comparison of data from an observing campaign over summer
    2018.  The power spectral density has been smoothed with
    a~\SI{50}{\micro\hertz} moving mean to aid clarity.}
  \label{fig:2018_comparison}
\end{figure}

\begin{table}
    \centering
    \caption{Velocity-calibrated white noise level and noise
      equivalent velocity (NEV) for four BiSON spectrophotometers, and
      comparison with the miniature prototype instrument.}
    \label{table:2018_performance}

    \begin{tabular}{l S[table-format=4.1] S[table-format=4.1] S[table-format=4.1]}

    \toprule

    Site & {FOM} & {Background Noise}          & {Noise Equivalent Velocity}\\
         &       & \si{\mps\squared\per\hertz} & \si{\centi\meter\per\second}~RMS\\

    \midrule

           Mount~Wilson &    82.6 &     1.6 &    14.2\\
             Sutherland &    42.9 &     4.3 &    23.1\\ 
      \textbf{BiSON:NG} &    19.6 &     9.9 &    35.2\\
               Narrabri &    27.9 &    10.4 &    36.1\\
           Las~Campanas &    25.0 &    13.5 &    41.1\\
                 Mark-I &     5.5 &    61.6 &    87.7\\
    %          Carnarvon & {-----} & {-----} & {-----}\\

    \bottomrule

    \end{tabular}

\end{table}

The prototype system was trialled at Iza{\~n}a, Tenerife, over two
site visits in~2017.  Light from the c{\oe}lostat supplying Mark-I was
shared with the new system.  The instrumentation continued to run
alongside Mark-I capturing contemporaneous data through summer~2018.

Figure~\ref{fig:2018_comparison} shows the power spectral density of
concatenated time series from each BiSON site for the summer 2018
observing campaign.  Specifically, data from 2018~April~14 to
2018~August~26 were selected since this is the period where the
secondary c{\oe}lostat mirror at Iza\~na is mounted in the summer
position.  A~\SI{50}{\micro\hertz} moving mean has been applied to
smooth the results and aid clarity.

Table~\ref{table:2018_performance} provides a summary of the
performance statistics from each site.  The ``FOM'' is a
signal-to-noise figure-of-merit.  The signal is estimated as the mean
power in the~\SIrange{2}{5}{\milli\hertz} band, and the noise
background as the mean power in the
upper~\SIrange{5.5}{12.5}{\milli\hertz} band.  This is not perfect
since there is, of course, noise within the signal band, and the upper
signal cut-off is not precise.  However, this definition has become a
network standard metric and so it is used here for consistency and
comparison.  The prototype instrument achieves a background noise
level of~\SI{9.9}{\mps\squared\per\hertz}, comparable to a noise
equivalent velocity RMS of~\SI{35.2}{\centi\meter\per\second}, and
this is similar performance to existing full-size BiSON
instrumentation.

The original Mark-I instrument has a much higher background noise
level due to slower polarisation switching rate, and so it experiences
much higher scintillation noise.  The poorer low-frequency performance
is due to the guiding accuracy of the c{\oe}lostat.  The new fibre-fed
instrument follows a similar low-frequency profile due to difficulty
maintaining the solar image within the fibre core.  In parallel with
work on the new instrumentation, the spectrophotometer at Mount Wilson
was also converted to fibre and is now the best performing in the
network, indicated by the red line with the lowest background white
noise level in Figure~\ref{fig:2018_comparison}.  This can be
considered an example of what can be achieved when the noise reduction
techniques applied here are combined with fast polarisation switching,
eliminating much of the atmospheric scintillation.

%%%%%%%%%%%%%%%%%%%%%%%%%%%%%%%%%%%%%%%%%%%%%%%%%%%%%%%%%%%%%%%%%%%%%%%%%%%%%%%%

% -*- coding: utf-8 -*-
%
% CONCLUSIONS.TEX
%
%   Steven Hale
%   2022 February 11
%   Birmingham, UK
%
% Conclusions
%

%%%%%%%%%%%%%%%%%%%%%%%%%%%%%%%%%%%%%%%%%%%%%%%%%%%%%%%%%%%%%%%%%%%%%%%%%%%%%%%%

\section{Conclusions}

The aim was to develop a miniaturised spectrophotometer that required
only a small amateur telescope mount, and used off-the-shelf
components where possible, in order to simplify the design and
maintenance requirements but without sacrificing performance.  We have
shown that the prototype instrument fulfils these aims, and produces
data of equal quality that can be seamlessly integrated into the
existing network.

Projects such as BiSON, and the extended observational network
potential of BiSON:NG, provide the necessary data that probes the
structure and rotation of the solar core, and continues to improve our
knowledge of the dynamo that drives the solar cycle.  Observation of
granulation noise at different heights in the solar atmosphere, probed
via different working points on the potassium~D1 absorption line, is
largely incoherent, and frequency regions dominated by oscillations
are almost fully coherent~\citep{2017MNRAS.472.3256L}.
Contemporaneous observation from many sites allows exploitation of
incoherence in granulation noise to beat down the observed noise level
through weighted averaging of multi-working-point
data~\citep{doi:10.1093/mnras/stu803}.  Such long term synoptic
observations of the Sun are important for improving detection of very
low-frequency modes and hence constraints on the structure of the deep
solar interior.

Following a period of reduced solar activity, it is important that
long-term whole-Sun seismic observations continue as we progress
through Cycle~\num{25} and into Cycle~\num{26} in order to obtain the
necessary data required to improve models of the solar dynamo.
Determining the accuracy of models and predictions is also important
for assessing the long-term impact on space weather.  Refreshing and
extending the existing ageing hardware will secure the ongoing
synoptic programme of helioseismic observations of whole-Sun
oscillations.

%%%%%%%%%%%%%%%%%%%%%%%%%%%%%%%%%%%%%%%%%%%%%%%%%%%%%%%%%%%%%%%%%%%%%%%%%%%%%%%%

% -*- coding: utf-8 -*-
%
% ACKNOWLEGEMENTS.TEX
%
%   Steven Hale
%   2022 February 11
%   Birmingham, UK
%
% Acknowledgements
%

%%%%%%%%%%%%%%%%%%%%%%%%%%%%%%%%%%%%%%%%%%%%%%%%%%%%%%%%%%%%%%%%%%%%%%%%%%%%%%%%

\section*{Acknowledgements}

We would like to thank all those who have been associated with BiSON
over the years.  We particularly acknowledge the invaluable technical
assistance at all our remote network sites.  During commissioning the
new prototype at Iza\~na: Dr.\,Pere {Pall{\'e}}, Dr.\,Teo {Roca
  Cort{\'e}s}, Antonio Pimienta, Santiago L\'{o}pez, and all the staff
at the Instituto de Astrof\'{i}sica de Canarias who have contributed
to running the {Mark-I} instrument over many years.  During testing of
the optical fibre light feed at Mount~Wilson:
Prof.\,Ed\,J.\,{Rhodes},\,Jr., and all former and current members of
the team of {USC} undergraduate observing assistants.  {BiSON} is
funded by the Science and Technology Facilities Council ({STFC}) grant
ST/V000500/1.

%%%%%%%%%%%%%%%%%%%%%%%%%%%%%%%%%%%%%%%%%%%%%%%%%%%%%%%%%%%%%%%%%%%%%%%%%%%%%%%%

% -*- coding: utf-8 -*-
%
% DATA.TEX
%
%   Steven Hale
%   2022 February 11
%   Birmingham, UK
%
% Data Availability
%

%%%%%%%%%%%%%%%%%%%%%%%%%%%%%%%%%%%%%%%%%%%%%%%%%%%%%%%%%%%%%%%%%%%%%%%%%%%%%%%%

\section*{Data Availability}

All data are freely available from the {BiSON} Open Data
Portal~--~\url{http://bison.ph.bham.ac.uk/opendata}.  Data products
are in the form of calibrated velocity residuals, concatenated into a
single time series from all {BiSON} sites.  Individual days of raw or
calibrated data, and also bespoke products produced from requested
time periods and sites, are available by contacting the authors.
Oscillation mode frequencies and amplitudes are available
from~\cite{2009MNRAS.396L.100B} and~\cite{2014MNRAS.439.2025D}.
Source files for this document are available from the GitLab
repository~\citep{gitlab}.

%%%%%%%%%%%%%%%%%%%%%%%%%%%%%%%%%%%%%%%%%%%%%%%%%%%%%%%%%%%%%%%%%%%%%%%%%%%%%%%%

%%%%%%%%%%%%%%%%%%%% REFERENCES %%%%%%%%%%%%%%%%%%

\bibliographystyle{rasti}
\bibliography{references}

%%%%%%%%%%%%%%%%% APPENDICES %%%%%%%%%%%%%%%%%%%%%

%\appendix
%\input{appendix}

%%%%%%%%%%%%%%%%%%%%%%%%%%%%%%%%%%%%%%%%%%%%%%%%%%

% Don't change these lines
\bsp	% typesetting comment
\label{lastpage}
\end{document}